\documentclass[showpacs,preprintnumbers,amsmath,amssymb,floatfix]{revtex4}

\usepackage{epsfig}
\usepackage{graphicx}% Include figure files
\usepackage{dcolumn}% Align table columns on decimal point
\usepackage{bm}% bold math
\usepackage{amsmath}
\usepackage{slashbox}
\usepackage{multirow}
\usepackage{color}
\usepackage{mathrsfs}

\begin{document}

\newcommand{\MSbar}{\overline{\rm MS}}
\def\be{\begin{equation}}
\def\ee{\end{equation}}
\def\bea{\begin{eqnarray}}
\def\eea{\end{eqnarray}}
\newcounter{Lcount}
\def\bl{\setcounter{Lcount}{0}
\begin{list}{\arabic{Lcount}.}{\usecounter{Lcount}\setlength{\leftmargin}{0.4cm}}}
\def\el{\end{list}}

%%%%%%%%%%%% Begin Cover Page %%%%%%%%%%%%%%%%%%%%%%%%%%%%%%%%%%%%%%%%%%

\preprint{DESY~11--142\hspace{13.95cm} ISSN 0418--9833}
\preprint{August 2011\hspace{16.55cm}}
\title{Timelike Single-logarithm-resummed Splitting Functions}
\author{S.\ Albino, P.\ Bolzoni, B.A.\ Kniehl}
%\footnote{%
%On leave of absence from Bogoliubov Laboratory of Theoretical Physics,
%Joint Institute for Nuclear Research, 141980 Dubna, Russia.}
\affiliation{{II.} Institut f\"ur Theretische Physik, Universit\"at Hamburg,\\
             Luruper Chaussee 149, 22761 Hamburg, Germany}
%\author{P.\ Bolzoni}
%\affiliation{{II.} Institut f\"ur Theoretische Physik, Universit\"at Hamburg,\\
%             Luruper Chaussee 149, 22761 Hamburg, Germany}
%\author{B.\ A.\ Kniehl}
%\affiliation{{II.} Institut f\"ur Theoretische Physik, Universit\"at Hamburg,\\
%             Luruper Chaussee 149, 22761 Hamburg, Germany}
\author{A.V.\ Kotikov}
\affiliation{{II.} Institut f\"ur Theoretische Physik, Universit\"at Hamburg,\\
             Luruper Chaussee 149, 22761 Hamburg, Germany}
\affiliation{Bogoliubov Laboratory of Theoretical Physics,
Joint Institute for Nuclear Research, 141980 Dubna, Russia}

\date{\today}
\begin{abstract}
We calculate the single logarithmic contributions to the quark singlet
and gluon matrix of 
timelike splitting functions at all orders in the modified minimal-subtraction
($\overline{\rm MS}$) scheme.
We fix two of the degrees of freedom of this matrix from the analogous results in the massive-gluon regularization scheme by using the relation between
that scheme and the $\overline{\rm MS}$ scheme.
We determine this scheme transformation from
the double logarithmic contributions to the timelike splitting functions and the
coefficient functions of inclusive particle production in $e^+ e^-$ annihilation now available in both schemes.
This approach fixes two of the four degrees of freedom, and a third degree of freedom is fixed by reasonable physical assumptions.
The results agree with the fixed-order results at next-to-next-to-leading order
in the literature.
\end{abstract}


\maketitle

%%%%%%%%%%%% End of Cover Page %%%%%%%%%%%%%%%%%%%%%%%%%%%%%%%%%%%%%%%%%

%%%%%%%%%%%%%% Begin Section I %%%%%%%%%%%%%%%%%%%%%%%%%%%%%%%%%%%%%%%%%
\section{Introduction}
\label{Intro}

According to the factorization theorem, the cross section ${\mathcal F}^h(x,Q^2)=Q^2 (d\sigma^h/dx)(x,Q^2)$ for the inclusive production of a hadron $h$
carrying a fraction $x$ of the available energy in a process with an energy scale $Q$ much greater than the asymptotic scale parameter $\Lambda_{\rm QCD}$ of QCD
can be calculated by the convolution
\be
{\mathcal F}^h(x,Q^2) = \sum_\alpha \int^1_x \frac{dz}{z} F_\alpha\left(z,\frac{\mu_f^2}{Q^2},a_s(\mu_f^2)\right)D^h_\alpha\left(\frac{x}{z},\mu_f^2\right),
\label{XSfromFFs}
\ee
where $F_\alpha\left(z,\mu_f^2/Q^2,a_s(\mu_f^2)\right)=Q^2 (d\sigma_\alpha/dz)\left(z,Q^2,\mu_f^2\right)$ 
is the equivalent process-dependent factorized cross section for the production
of a parton $\alpha$ carrying away a fraction $z$ of the available energy, 
which contains all subprocesses with energy scale greater than the arbitrary factorization scale $\mu_f$,
$D^h_\alpha\left(z,\mu_f^2\right)$ is the factorized fragmentation function (FF) for the fragmentation of a parton $\alpha$ to a hadron $h$ carrying away
a fraction $z$ of the energy of this parton, 
which contains all subprocesses with energy scale less than $\mu_f$, and
$a_s=\alpha_s/(2\pi)$, with $\alpha_s$ being the strong-coupling constant.
These partonic cross sections are perturbatively calculable. I.e.\ the series
\be
F_\alpha\left(z,\frac{\mu_f^2}{Q^2},a_s(\mu_f^2)\right)=\sum_{n=n_0}^\infty a_s^n(\mu_f^2) F_\alpha^{(n)}\left(z,\frac{\mu_f^2}{Q^2}\right)
\label{pertexpanofFalpha}
\ee
approximates $F_\alpha\left(z,\mu_f^2/Q^2,a_s(\mu_f^2)\right)$ for sufficiently small
values of $a_s$.
We refer to this approach to calculations, namely expanding in $a_s$ with coefficients that are independent of $a_s$,
as the {\it fixed-order} (FO) approach.
The scale $Q$ will be loosely defined to be the scale which $\mu_f$
should be chosen to have the same order of magnitude as in order that the perturbative series for 
$F_\alpha\left(z,\mu_f^2/Q^2,a_s(\mu_f^2)\right)$ be as convergent as possible.
To be explicit, this is because the coefficients $F_\alpha^{(n)}\left(z,\mu_f^2/Q^2\right)$ in Eq.\ (\ref{pertexpanofFalpha})
grow like $\ln^{n+p}(\mu_f^2/Q^2)$ as $\mu_f^2/Q^2\rightarrow \infty$, where $p$ is an integer that is independent of $n$.
The predictive power of this approach follows from the facts that the FFs are process independent
and the dependence of the FFs on $\mu_f$ obeys the 
Dokshitzer-Gribov-Lipatov-Altarelly-Parisi (DGLAP) evolution equation
\cite{Gribov:1972ri,Gribov:1972rt,Altarelli:1977zs,Dokshitzer:1977sg},
\be
\frac{d}{d\ln \mu_f^2}D_\alpha^h(z,\mu_f^2)=\sum_\beta \int^1_z \frac{dz}{z} P_{\alpha \beta}\left(z,a_s(\mu_f^2)\right)D_\beta^h\left(\frac{x}{z},\mu_f^2\right),
\label{DGLAPxspace}
\ee
where $P_{\alpha \beta}(z,a_s)$ are the $\alpha\to\beta$ splitting functions, which
are perturbatively calculable in the FO approach for sufficiently small values of $a_s$, the perturbative series taking the form
\be
P_{\alpha\beta}\left(z,a_s\right)=\sum_{n=1}^\infty a_s^n P_{\alpha\beta}^{(n-1)}(z).
\label{expanP}
\ee
%Here we have used matrix notation, i.e.\ $P$ is the matrix with components $P_{\alpha \beta}$.

However, the FO approach fails when $x$ is too small, due to the presence of unresummed large soft-gluon logarithms (SGLs) 
in the timelike splitting functions and in the hard partonic cross sections.
This means that small-$x$ measurements cannot be used to provide constraints on FFs at small values of $z$.
They also cannot be used to improve the FFs at higher values of $z$ because, according to Eq.\ (\ref{XSfromFFs}),
the cross section at $x$ depends on the FFs $D^h_\alpha\left(z,\mu_f^2\right)$ at all $z$ values in the range $x\leq z\leq 1$.

To improve the accuracy at small values of $x$, the SGLs of each class appearing in the FO expressions must be determined to all orders.
The {\it double logarithms} (DLs), being the largest SGLs, 
are known to all orders in the $\overline{\rm MS}$ scheme
for the timelike splitting functions \cite{Albino:2005gd} and the coefficient functions 
for inclusive hadron production in $e^+ e^-$ annihilation \cite{Albino:2011si,Albino:2011bf}.
The {\it single logarithms} (SLs) in the splitting functions to all orders 
are known only in the massive-gluon regularization (MG) scheme \cite{Mueller:1983cq}.
Because the FO approach and the resummed SGLs can be consistently combined 
as discussed in Refs.\  \cite{Albino:2005gd,Albino:2005gg,Albino:2005cq} to give an approach which can describe
data from the smallest to the largest values of $x$
and because FO calculations in the $\overline{\rm MS}$ scheme are known to
next-to-leading order (NLO) and beyond, while those in the MG scheme are usually not,
it is necessary to determine the SLs in the $\overline{\rm MS}$ splitting functions.
Furthermore, FFs are usually determined in the $\overline{\rm MS}$ scheme.

In this paper,
we first partially constrain the complete SL contributions to the $\overline{\rm MS}$ splitting functions using three key ingredients: 
firstly, the DL contributions to the splitting functions in these two schemes;
secondly, the SL contributions to the MG splitting functions; and 
thirdly, the DL contribution to the scheme change between the MG and $\overline{\rm MS}$ schemes.
The third ingredient can be obtained because the DL contribution to the gluon coefficient function 
of $e^+ e^-$ annihilation is known in the MG scheme
and we recently calculated the same quantity in the $\overline{\rm MS}$ scheme \cite{Albino:2011si,Albino:2011bf}.
To completely constrain the SL contributions to the $\overline{\rm MS}$ splitting functions,
we then introduce some reasonable assumptions that fix the next-to-lowest order of the scheme change:
We demand that our results are consistent with the next-to-next-to-leading-order
(NNLO) splitting functions, and also that the 
matrix exhibits certain non-singular properties at small values of $z$.

This paper is organised as follows.
In section \ref{Gfihpcs}, we discuss the calculations of factorized cross sections in general.
In section \ref{Sgl}, we introduce SGLs and present the DLs in the MG and $\overline{\rm MS}$ schemes for the coefficient functions in $e^+ e^-$ annihilation
and for the timelike splitting functions. 
We formalize the relation between calculations in different schemes in section \ref{Gsc}.
In section \ref{Slitsf}, we use these results together with the
SLs in the timelike splitting functions in the MG scheme determined in Ref.\ \cite{Mueller:1983cq} to
determine the SLs for combinations of the splitting functions in the $\overline{\rm MS}$ scheme.
Finally, we present our conclusions in section \ref{conc}.

\section{General factorized inclusive hadron production cross sections}
\label{Gfihpcs}

In this section, we consider the general structure of the calculations of factorized cross sections that will be useful later.
We will find it convenient to work in Mellin space, defined by the (invertible) Mellin transform
\be
f(\omega)=\int_0^1 dx\, x^\omega f(x),
\ee
because $x$-space convolutions reduce to simple products.
In particular, removing the superscript $h$ from now on,
Eqs.\ (\ref{XSfromFFs}), (\ref{DGLAPxspace}), and (\ref{expanP}) respectively
become
\begin{eqnarray}
{\mathcal F}(\omega,Q^2)&=&\sum_\alpha F_\alpha\left(\omega,\frac{\mu_f^2}{Q^2},a_s(\mu_f^2)\right)D_\alpha(\omega,\mu_f^2),
\label{XSfromFFsMell}
\\
\frac{d}{d\ln \mu_f^2}D_\alpha(\omega,\mu_f^2)&=&\sum_\beta P_{\alpha \beta}\left(\omega,a_s(\mu_f^2)\right)D_\beta\left(\omega,\mu_f^2\right),
\label{DGLAPMell}
\\
 P_{\alpha\beta}(\omega,a_s)
&=&\sum_{n=1}^\infty a_s^n P_{\alpha\beta}^{(n-1)}(\omega).
\label{expanPomega}
\end{eqnarray}
%are the Mellin tranform of the corresponding kernels $P^{(n-1)}(z)$ in 
%Eq.\ (\ref{expanP}).

According to Eq.\ (\ref{XSfromFFsMell}), the cross section is invariant under any change of {\it parton basis} $F_\alpha \rightarrow \overline{F}_\alpha = \sum_\beta F_\beta (Y^{-1})_{\beta \alpha}$
and $D_\alpha \rightarrow \overline{D}_\alpha = \sum_\beta Y_{\alpha \beta} D_\beta$, where $Y$ is any invertible matrix
which is independent of $\omega$, $\mu_f^2$, and $Q^2$.
In matrix notation, $\overline{F}=FY^{-1}$ and $\overline{D} = Y D$.
For example, the SU$(n_f)$ symmetry of the DGLAP equation in the $\overline{\rm MS}$ scheme for $n_f$ active flavours of quarks
and the charge conjugation symmetry of QCD
imply that $P$ is reduced to block-diagonal form when the parton basis is chosen such that the FFs consist of the quark singlet component,
\be
D_\Sigma = \frac{1}{n_f}\sum_{J=1}^{n_f} \left(D_{q_{J}} +D_{\bar{q}_J}\right),
\ee
with $q_{J}$ ($\bar{q}_J$) being the (anti)quark of flavour $J$,
the quark non-singlet component,
\be
D_{q_J,{\rm NS}}=D_{q_J}+D_{\bar{q}_J}-D_\Sigma,
\ee
the valence-quark singlet and non-singlet components, and the gluon component, $D_g$.
In this basis, for 
\begin{eqnarray}
D=\left(\begin{array}{c}D_\Sigma\\D_g\end{array}\right)
\label{Sigandg2comp}
\end{eqnarray}
in Eq.\ (\ref{DGLAPMell}), we have the 2$\times$2 matrix
\begin{eqnarray}
P=\left(\begin{array}{cc}P_{\Sigma \Sigma} & P_{\Sigma g} \\ P_{g \Sigma} & P_{gg}\end{array}\right),
\label{defofPintermsofPSigg}
\end{eqnarray}
while, for $D=D_{q_J,{\rm NS}}$, we have the single flavour-independent quantity $P=P_{\rm NS}$, and simlarly for the valence-quark singlet and non-singlets.

An alternative basis, which is used in some applications and will be needed
later, is that in which the LO splitting function matrix is diagonal, i.e.
\begin{eqnarray}
D=\left(\begin{array}{c}D_+\\D_-\end{array}\right)
\end{eqnarray}
and
\begin{eqnarray}
P=\left(\begin{array}{cc}P_{++} & P_{+-} \\ P_{-+} & P_{--}\end{array}\right),
\end{eqnarray}
where, defining the projectors $\alpha$, $\beta$, and
$\epsilon$ by \cite{Buras:1979yt}
\be
\alpha = \frac{P^{(0)}_{\Sigma \Sigma}-P^{(0)}_{++}}{P^{(0)}_{--}
-P^{(0)}_{++}},\qquad
\beta = \frac{P^{(0)}_{g \Sigma}}{P^{(0)}_{--}-P^{(0)}_{++}},\qquad
\epsilon = \frac{P^{(0)}_{\Sigma g} }{P^{(0)}_{--}-P^{(0)}_{++}},
\label{1.4}
\ee
we have
\be
\begin{split}
D_{+} &= (1-\alpha)D_\Sigma - \beta D_g,\\
D_{-} &= \alpha D_\Sigma + \beta D_g,
\label{DminusfromDSDg}
\end{split}
\ee
and, for all $k\geq 0$,
\bea
P^{(k)}_{--} &=& \alpha  P^{(k)}_{\Sigma \Sigma} 
+ \beta  P^{(k)}_{\Sigma g} + \epsilon  P^{(k)}_{g \Sigma}
+ (1-\alpha)  P^{(k)}_{gg}, \nonumber \\
 P^{(k)}_{+-} &=& P^{(k)}_{--} - \left(P^{(k)}_{\Sigma \Sigma} +
\frac{1-\alpha}{\epsilon}  P^{(k)}_{\Sigma g} \right), \nonumber \\
P^{(k)}_{++} &=&  P^{(k)}_{\Sigma \Sigma} +  P^{(k)}_{gg} - P^{(k)}_{--},
\nonumber \\
 P^{(k)}_{-+} &=& P^{(k)}_{++} - \left(P^{(k)}_{\Sigma \Sigma} -
\frac{\alpha}{\epsilon}  P^{(k)}_{\Sigma g} \right) ~=~  P^{(k)}_{gg} - 
\left(P^{(k)}_{--} - \frac{\alpha}{\epsilon}  P^{(k)}_{\Sigma g} \right).
\label{1.7}
\eea
Note, of course, that $P^{(0)}_{\pm\mp} = 0$ by definition.
In one important simplification of QCD, namely ${\mathcal N}=4$ super Yang-Mills theory, this basis is actually more natural than the basis of 
quark singlet and gluon because the diagonal splitting functions $P^{(k)}_{\pm\pm}$ can be expressed 
in all orders of perturbation theory as one universal function with shifted arguments \cite{Kotikov:2002ab}.

In general, because both the Mellin transform and the change of parton basis are invertible,
we will not specify whether the $x$-space convolution of two $x$-space functions or the product of their Mellin transforms is being calculated,
nor which parton basis is being used, nor whether only a subspace of the full parton space (achieved by setting combinations 
of FFs to zero) is being considered,
but simply write Eqs.\ (\ref{XSfromFFs}) and (\ref{XSfromFFsMell}) as
\be
{\mathcal F}=F D,
\label{FhFDh}
\ee
and Eqs.\ (\ref{DGLAPxspace}) and (\ref{DGLAPMell}) as
\be
\frac{d}{d\ln \mu_f^2}D=PD.
\label{DGLAP}
\ee

Inclusive particle production in $e^+ e^-$ annihilation provides a simple example of this formalism. 
In this case, $Q$ is conveniently chosen to be the c.m.\ energy,
and the cross section takes the form
\be
{\mathcal F} = \sum_{J=1}^{n_f} F_{q_J,{\rm NS}} D_{q_J,{\rm NS}}+F_\Sigma D_\Sigma +F_g D_g,
\label{Fhinepem}
\ee
where
\begin{eqnarray}
F_{q_J,{\rm NS}}\left(\omega,\frac{\mu_f^2}{Q^2},a_s(\mu_f^2)\right)&=&
Q^2 \sigma_0(Q^2)N_c Q_{q_J}(Q^2) C_{\rm NS}\left(\omega,\frac{\mu_f^2}{Q^2},a_s(\mu_f^2)\right),
\nonumber\\
F_\Sigma\left(\omega,\frac{\mu_f^2}{Q^2},a_s(\mu_f^2)\right)&=&
Q^2 \sigma_0(Q^2)N_c n_f \langle Q(Q^2) \rangle C_\Sigma\left(\omega,\frac{\mu_f^2}{Q^2},a_s(\mu_f^2)\right),
\nonumber\\
F_g\left(\omega,\frac{\mu_f^2}{Q^2},a_s(\mu_f^2)\right)&=&
Q^2 \sigma_0(Q^2)N_c n_f \langle Q(Q^2) \rangle C_g\left(\omega,\frac{\mu_f^2}{Q^2},a_s(\mu_f^2)\right),
\label{eq16}
\end{eqnarray}
with $\sigma_0(Q^2)$ being the lowest-order (LO) cross section 
for the process $e^+ e^-\rightarrow \mu^+\mu^-$, $N_c$ the number of quark colours in QCD, $Q_{q_J}(Q^2)$ the effective electroweak charge of quark $q_J$, and
$\langle Q(Q^2) \rangle=\sum_{J=1}^{n_f}Q_{q_J}(Q^2)/n_f$.
Note that Eq.\ (\ref{eq16}) is, strictly speaking, dependent on $Q$ through $M_Z^2/Q^2$ (in $Q_{q_J}(Q^2)$) as well as through
$\mu_f^2/Q^2$, but these dependences are not shown for brevity.
The coefficient functions $C_X$ ($X={\rm NS},\Sigma,g$) in the FO approach in Mellin space may be found, e.g., in Ref.~\cite{Albino:2008gy}.
%take the form
%\begin{eqnarray}
%C_{\rm NS}\left(\omega,\frac{\mu_f^2}{Q^2},a_s\right)&=&1+O(a_s),
%\nonumber\\
%C_\Sigma\left(\omega,\frac{\mu_f^2}{Q^2},a_s\right)&=&
%C_{\rm NS}\left(\omega,\frac{\mu_f^2}{Q^2},a_s\right)+O(a_s^2),
%\nonumber\\
%C_g\left(\omega,\frac{\mu_f^2}{Q^2},a_s\right)&=&O(a_s).
%\end{eqnarray}
It will be convenient later to write
\be
{\mathcal F}=Q^2\sigma_0(Q^2)N_c n_f \langle Q(Q^2) \rangle CD.
\ee
For example, for the quark singlet and gluon contribution in Eq.\ (\ref{Fhinepem}), 
\be
{\mathcal F}=F_\Sigma D_\Sigma +F_g D_g,
\label{Finsingletcase}
\ee
$D$ is given by Eq.\ (\ref{Sigandg2comp}), and
\be
C=(C_\Sigma,C_g).
\ee
We will set $\mu_f=Q$ for simplicity, in which case it is convenient to define
\be
C_X(\omega,a_s)=C_X(\omega,1,a_s)\qquad(X={\rm NS},\Sigma,g).
\ee

\section{Soft-gluon logarithms}
\label{Sgl}

Since the non-singlet inclusive partonic production cross sections $F_{q_J,{\rm NS}}$ and the non-singlet splitting functions are free of SGLs, 
they do not concern us, and so we will not discuss them further. 
From now on, inclusive particle production cross sections will be assumed to take the form in Eq.\ (\ref{Finsingletcase}).
The inclusive partonic production cross sections $F$ calculated in the FO approach may exhibit a singular behaviour in Mellin space as $\omega \rightarrow 0$.
This is caused by SGLs, which grow like $1/\omega^p$ for $p\ge 1$.
In $x$ space, these SGLs take the form of quantities that grow like $\ln^{p-1}x$ as $x\rightarrow 0$.
Such strong singularities are non-physical and become weaker or even disappear after being resummed to all orders.
The resummed SGLs in $F$ take the form of the series
\be
F=\sum_{m=0}^\infty \left(\frac{a_s}{\omega}\right)^m F^{[m]}\left(\frac{a_s}{\omega^2}\right).
\label{FSGL}
\ee
For such a series to converge, at least asymptotically, it is necessary that $a_s \ll 1$ and $\omega=O(\sqrt{a_s})$.
The DLs, namely those SGLs for which $m=0$ in Eq.\ (\ref{FSGL}), of the inclusive partonic production cross sections
for $e^+ e^-$ annihilation in the $\overline{\rm MS}$ scheme, when $D$ is given by Eq.\ (\ref{Sigandg2comp}), 
take the form \cite{Albino:2011si,Albino:2011bf}
\be
C=(1,C_g^{\rm DL}),
\label{CbarSigmagDL}
\ee
where
\be
C_g^{\rm DL}(\omega,a_s)=\frac{2C_F}{C_A}\left[\sqrt{\frac{\omega}
{4\gamma(\omega,a_s)+\omega}}-1\right],
\ee
with
\be
\gamma(\omega,a_s) = \frac{1}{4}(-\omega+\sqrt{\omega^2+16C_A a_s})
\label{gammaDL},
\ee
and $C_{\rm NS}=1$.
They were also determined in Ref.\ \cite{Mueller:1982cq} in the MG scheme, indicated in this paper by an overline, to be
\be
\overline{C}=(1,\overline{C}_g^{\rm DL}),
\label{CSigmagDL}
\ee
where
\be
\overline{C}_g^{\rm DL}(\omega,a_s)=\frac{C_F}{C_A}\left[\frac{\omega}
{4\gamma(\omega,a_s)+\omega}-1\right],
\ee
and $\overline{C}_{\rm NS}=1$.

The resummation of the SGLs in $P$ take the form of the series
\be
P=\sum_{m=1}^\infty \left(\frac{a_s}{\omega}\right)^m P^{[m-1]}\left(\frac{a_s}{\omega^2}\right).
\label{SGLexpanofP}
\ee
The full DL contribution to $P$, namely the SGLs for which $m=1$ in Eq.\ (\ref{SGLexpanofP}), will be written as
$P^{\rm DL}=(a_s/\omega)P^{[0]}(a_s/\omega^2)$.
When $D$ is given by Eq.\ (\ref{Sigandg2comp}), it is given in the $\overline{\rm MS}$ scheme by
\be
P^{\rm DL}(\omega,a_s)=A\gamma(\omega,a_s),
\label{PDLSigg}
\ee
where $\gamma$ is given in Eq.\ (\ref{gammaDL}) and
\be
A=\left(\begin{array}{cc} 0 & \frac{2C_F}{C_A} \\ 0 & 1\end{array}\right),
\ee
which obeys the projection operator property $A^2=A$.
For the quark non-singlets, $\overline{P}^{\rm DL}=0$.
The DLs in $P$ in the MG scheme are the same as those in the $\overline{\rm MS}$ scheme, 
i.e., when $D$ is given by Eq.\ (\ref{Sigandg2comp}),
\be
\overline{P}^{\rm DL}(\omega,a_s)=A\gamma(\omega,a_s),
\label{PbarDLSigg}
\ee
and $\overline{P}^{\rm DL}=0$ for the quark non-singlets.

\section{General scheme changes}
\label{Gsc}

Results in one scheme, such as the splitting functions in the $\overline{\rm MS}$ scheme,
may be obtained from the analogous results in another scheme, such as the MG scheme, 
once the relation between the two schemes is known to the appropriate accuracy.
To obtain the form of this relation,
let $F$ ($D$) and $\overline{F}$ ($\overline{D}$) be respectively the partonic cross sections (FFs) in any two different schemes.
Since the cross section is scheme independent, then as well as Eq.\ (\ref{FhFDh}) we have ${\mathcal F} =\overline{F} \overline{D}$. 
Comparing this last result with Eq.\ (\ref{FhFDh}) and treating
$D_\alpha(z,\mu_f^2)$ as arbitrary functions,
we find that
\begin{eqnarray}
F&=&\overline{F}Z,
\label{schemechangeF}
\\
D&=&Z^{-1}\overline{D},
\label{schemechangeD}
\end{eqnarray}
where, in Mellin space, $Z$ is an invertible matrix that depends on $\omega$ and $\mu_f^2$.
Note, therefore, that Eqs.\ (\ref{schemechangeF}) and (\ref{schemechangeD}) 
are generalizations of the change of parton basis considered just after Eq.\ (\ref{DGLAPMell}).
The (matrix of) splitting function(s) $\overline{P}$ is defined to be that which appears in the DGLAP equation
in the new scheme, which emerges from Eq.\ (\ref{DGLAP}) by substituting
$D$ and $P$ with $\overline{D}$ and $\overline{P}$, respectively.
%, Eq.\ (\ref{DGLAP}), for the new scheme, i.e.\
%\be
%\frac{d}{d\ln \mu_f^2}\overline{D}=\overline{P}\overline{D}.
%\ee
Then, it follows from Eqs.\ (\ref{DGLAP}) and (\ref{schemechangeD}) that the
relation between the splitting functions in two different schemes is given by
\be
P=Z^{-1}\overline{P}Z-Z^{-1}\frac{dZ}{d\ln \mu_f^2}.
\label{reloverlinePtoP}
\ee

Now, consider a general expansion of perturbatively calculable quantities, such as $F_\alpha$ and $P$, in some variable $x(\omega,a_s)$,
with coefficients that depend on $y(\omega,a_s)$, i.e.\
\begin{eqnarray}
F&=&\sum_{n=0}^\infty x^n F^{\{n\}}(y),
%\label{xyexpanF}
\nonumber
\\
P&=&\sum_{n=1}^\infty x^n P^{\{n-1\}}(y).
\label{xyexpanP}
\end{eqnarray}
For example, in the FO approach (Eqs.\ (\ref{pertexpanofFalpha}) and (\ref{expanPomega})), $x=a_s$ and $y=\omega$ in 
%Eqs.\ (\ref{xyexpanF}) and 
Eq.\ (\ref{xyexpanP}),
while in the SGL approach (Eqs.\ (\ref{FSGL}) and (\ref{SGLexpanofP})), $x=a_s/\omega$ and $y=a_s/\omega^2$ in 
%Eqs.\ (\ref{xyexpanF}) and 
Eq.\ (\ref{xyexpanP}) (excluding terms that are non-singular as $\omega \to 0$).
We restrict our schemes to be such that, if the perturbative series for $F$ begins at $O(x^n)$, 
the perturbative series for $\overline{F}$ also begins at $O(x^n)$.
Thus,
\be
Z(\omega,a_s)=\sum_{n=0}^\infty Z^{\{n\}}(y) x^n.
\label{genpertforZ}
\ee

Note that, in Eq.\ (\ref{reloverlinePtoP}), the first term $Z^{-1}\overline{P}Z=O(x)$ while the second term $Z^{-1}dZ/d\ln \mu_f^2=O(x^2)$.
Thus $P^{\{0\}}=Z^{\{0\}-1}\overline{P}^{\{0\}}Z^{\{0\}}$.
The result $P^{(0)}=\overline{P}^{(0)}$ no longer holds in general, 
but rather if and only if $Z^{\{0\}}$ commutes with $\overline{P}^{\{0\}}$. 
This is trivially the case in the FO approach
because the schemes used in the literature are (usually) such that $Z^{\{0\}}(y)=1$, i.e.\
\be
Z(\omega,a_s)=1+\sum_{n=1}^\infty Z^{(n)}(\omega) a_s^n.
\ee
However, in the SGL approach, where
\be
Z(\omega,a_s)=\sum_{m=0}^\infty Z^{[m]}\left(\frac{a_s}{\omega^2}\right) \left(\frac{a_s}{\omega}\right)^m,
\label{formofZinSGLapproach}
\ee
we must allow for the possibility that $Z^{[0]}(y)$ is any function of $y$.
We will see later that $Z^{[0]}$ does in fact commute with $\overline{P}^{[0]}$, at least for the MG and $\overline{\rm MS}$ schemes.

\section{Single logarithms in the splitting functions}
\label{Slitsf}

The SL contributions to the timelike splitting functions have already been calculated in the MG scheme \cite{Mueller:1982cq,
Mueller:1983js,Mueller:1983cq} and are given by 
\begin{eqnarray}
\overline{P}^{\rm SL}_{\Sigma \Sigma}&=&0,
\label{mgsplittingfunctions1}\\
\overline{P}^{\rm SL}_{\Sigma g}&=&
\frac{2C_F}{C_A}\left\{\left[\overline{P}^{\rm SL}_{g g}+\frac{1}{6}(11C_A+4 n_f T_R) a_s\right]
+\omega\left(\frac{1}{6}+\frac{1}{3}\frac{n_f T_R}{C_A}
-\frac{2}{3}\frac{C_F n_f T_R}{C_A^2}\right)\left(\gamma-\frac{2 a_s C_A}{\omega}\right)\right\}
-3C_F a_s,
\label{mgsplittingfunctions2}\\
\overline{P}^{\rm SL}_{g \Sigma}&=&\frac{2}{3} T_R n_f a_s,
\\
\overline{P}^{\rm SL}_{g g}&=&-\frac{1}{6}\frac{\omega^3 (11 C_A+4 n_f T_R)}{(4 \gamma+\omega)^3}a_s -\frac{2}{3} 
\frac{\omega [55 C_A^2 (2 \gamma + \omega) + 4 C_A n_f T_R (6 \gamma + 5 \omega) + 
    8 C_F n_f T_R\omega ]}
{(2 \gamma+\omega) (4 \gamma+\omega)^3}a_s^2
\nonumber\\
&&{}-\frac{16}{3}\frac{(11 C_A^3 + 12 C_A^2 n_f T_R + 16 C_A C_F n_f T_R)}{(2 \gamma + \omega) (4 \gamma +
\omega)^3}a_s^3,
\label{mgsplittingfunctions3}
\end{eqnarray}
where $\gamma=\gamma(\omega,a_s)$ is given by Eq.\ (\ref{gammaDL}).
In Eq.\ (\ref{mgsplittingfunctions2}),
we have taken the opportunity to correct some obvious typographical errors in
 Eq.\ (38) 
of Ref.\ \cite{Mueller:1983cq}\footnote{%
On the first line, $\gamma_{22}^{(0,0,0)GG}$ should be $\gamma_{22}^{(0,0,0)GF}$. 
On the third line, $\gamma_{11}^{(0,0,0)GG}$ should be $\gamma_{22}^{(0,0,0)GG}$. 
On the fourth line, the denominator of $\gamma_{11}^{(1,0,0)GG}$ should be 
$(n-1-2\gamma_n^{(0)})^2$.}.
It is the goal of this section to perform the scheme change given in Eq.\ (\ref{reloverlinePtoP}) 
on Eqs.\ (\ref{mgsplittingfunctions1})--(\ref{mgsplittingfunctions3}) in order to constrain and then to attempt to determine the SL contributions
in the $\overline{\rm MS}$ scheme.

We first calculate $Z^{[0]}$ from the DLs
in the coefficient and splitting functions: 
with the help of Eqs.\ (\ref{CbarSigmagDL}) and (\ref{CSigmagDL}), Eq.\ (\ref{schemechangeF}) becomes
\be
(1,C_g^{\rm DL})=(1,\overline{C}_g^{\rm DL})\left(
\begin{array}{cc}
Z^{[0]}_{\Sigma \Sigma} & Z^{[0]}_{\Sigma g}\\
Z^{[0]}_{g \Sigma} & Z^{[0]}_{gg}
\end{array}\right).
\ee
Using this result to eliminate $Z^{[0]}_{\Sigma \Sigma}$ and $Z^{[0]}_{\Sigma g}$ gives
\be
Z^{[0]}=\left(\begin{array}{cc}1-\overline{C}_g^{\rm DL}Z^{[0]}_{g \Sigma} & C_g^{\rm DL}-\overline{C}_g^{\rm DL}Z^{[0]}_{g g}\\
Z^{[0]}_{g \Sigma} & Z^{[0]}_{gg}\end{array}\right).
\ee
Next, we note that, because $dZ/d\ln \mu_f^2$ is free of DLs as discussed immediately after Eq.\ (\ref{genpertforZ})
\footnote{I.e.\ $Z^{-1}dZ/d\ln \mu_f^2=O((a_s/\omega)^2)$ with $Z$ taking the form in Eq.\ (\ref{formofZinSGLapproach}).}, the DLs in Eq.\ (\ref{reloverlinePtoP}) obey
\be
P^{\rm DL}=Z^{[0]-1} \overline{P}^{\rm DL}Z^{[0]}.
\ee
Using Eqs.\ (\ref{PDLSigg}) and (\ref{PbarDLSigg}), we find that 
\be
[Z^{[0]},A]=0,
\label{Z0commuteA}
\ee
i.e.\
\begin{eqnarray}
\left(\begin{array}{cc}0 & \frac{2C_F}{C_A}(1-\overline{C}_g^{\rm DL}Z^{[0]}_{g \Sigma})+C_g^{\rm DL}-\overline{C}_g^{\rm DL}Z^{[0]}_{gg}\\
0 & \frac{2C_F}{C_A}Z^{[0]}_{g \Sigma}+Z^{[0]}_{gg}\end{array}\right)=
\left(\begin{array}{cc}\frac{2C_F}{C_A}Z^{[0]}_{g \Sigma} & \frac{2C_F}{C_A}Z^{[0]}_{g g}\\
Z^{[0]}_{g \Sigma} & Z^{[0]}_{g g}
\end{array}\right),
\end{eqnarray}
from which we find that
\begin{eqnarray}
Z^{[0]}_{g \Sigma}&=&0,
\nonumber\\
Z^{[0]}_{g g}&=&\frac{\frac{2C_F}{C_A}+C_g^{\rm DL}}{\frac{2C_F}{C_A}+\overline{C}_g^{\rm DL}}.
\end{eqnarray}
Thus, finally,
\begin{eqnarray}
Z^{[0]}=\left(\begin{array}{cc}1 & \frac{C_g^{\rm DL}-\overline{C}_g^{\rm DL}}{1+\frac{C_A}{2C_{F}}\overline{C}_g^{\rm DL}}\\
0 & \frac{1+\frac{C_A}{2C_{F}}C_g^{\rm DL}}{1+\frac{C_A}{2C_{F}}\overline{C}_g^{\rm DL}}\end{array}\right).
\end{eqnarray}
We note that 
\begin{eqnarray}
Z^{[0]-1}=\left(\begin{array}{cc}1 & \frac{\overline{C}_g^{\rm DL}-C_g^{\rm DL}}{1+\frac{C_A}{2C_{F}}C_g^{\rm DL}}\\
0 & \frac{1+\frac{C_A}{2C_{F}}\overline{C}_g^{\rm DL}}{1+\frac{C_A}{2C_{F}}C_g^{\rm DL}}\end{array}\right)
\end{eqnarray}
is obtained from $Z^{[0]}$ by taking $C_g^{\rm DL}\leftrightarrow \overline{C}_g^{\rm DL}$, as expected.

We are now in a position to constrain the SLs in $P$.
Using Eq.\ (\ref{Z0commuteA}), the SLs (divided by $(a_s/\omega)^2$ for convenience) in the $\overline{\rm MS}$ splitting functions 
in Eq.\ (\ref{reloverlinePtoP}) are given by
\be
P^{[1]}=Z^{[0]-1} \overline{P}^{[1]}Z^{[0]}+\beta_0 \omega^2 Z^{[0]-1}\frac{dZ^{[0]}}{da_s}+R^{[1]},
\label{PSLfromPSLoverline}
\ee
where $\beta_0=(11/6)C_A-(2/3)T_R n_f$ is first coefficient in the series
$\beta(a_s)=-\sum_{n=0}^\infty \beta_n a_s^{n+2}$ that determines the running of the coupling via
$da_s(\mu^2)/d\ln \mu^2=\beta(a_s(\mu^2))$
and we have defined
\be
R^{[1]}=Z^{[0]-1}[\overline{P}^{[0]},Z^{[1]}]=
P_{gg}^{[0]}\left(\begin{array}{cc}
\frac{2C_F}{C_A}\frac{1}{Z_{gg}^{[0]}} Z_{g\Sigma}^{[1]} &
\frac{2C_F}{C_A}(Z_{gg}^{[1]}-Z_{\Sigma \Sigma}^{[1]})-Z_{\Sigma g}^{[1]}-\left(\frac{2C_F}{C_A}\right)^2\left(\frac{1}{Z_{gg}^{[0]}}-1\right) Z_{g\Sigma}^{[1]}\\
\frac{1}{Z_{gg}^{[0]}}Z_{g\Sigma}^{[1]} & -\frac{2C_F}{C_A}\,\frac{1}{Z_{gg}^{[0]}} Z_{g\Sigma}^{[1]}
\end{array}\right).
\label{defofR}
\ee
From Eq.\ (\ref{SGLexpanofP}) and the definitions that follow it, we have
$P_{gg}^{[0]}=\left(a_s/\omega\right)^{-1}\gamma(\omega,a_s)$.
Explicitly at SL accuracy, Eqs.\ (\ref{PSLfromPSLoverline}) and (\ref{defofR}) read 
\begin{eqnarray}
P^{\rm SL}_{\Sigma \Sigma}&=&-\frac{C_g^{\rm DL}-\overline{C}_g^{\rm DL}}{1+\frac{C_A}{2C_{F}}C_g^{\rm
DL}}\overline{P}^{\rm SL}_{g \Sigma}+R^{\rm SL}_{\Sigma \Sigma},
\nonumber\\
P^{\rm SL}_{\Sigma g}&=&\frac{1+\frac{C_A}{2C_{F}}C_g^{\rm DL}}{1+\frac{C_A}{2C_{F}}\overline{C}_g^{\rm DL}}\left(\overline{P}^{\rm SL}_{\Sigma g}
+\beta_0 a_s^2\frac{d}{da_s}\,\frac{C_g^{\rm DL}-\overline{C}_g^{\rm DL}}{1+\frac{C_A}{2C_{F}}C_g^{\rm DL}}\right)
-\frac{\left(\overline{C}_g^{\rm DL}-C_g^{\rm DL}\right)^2}{\left(1+\frac{C_A}{2C_{F}}C_g^{\rm DL}\right)\left(1+\frac{C_A}{2C_{F}}\overline{C}_g^{\rm DL}\right)} 
\overline{P}^{\rm SL}_{g \Sigma}
\nonumber\\
&&{}+\frac{C_g^{\rm DL}-\overline{C}_g^{\rm DL}}{1+\frac{C_A}{2C_{F}}\overline{C}_g^{\rm DL}}
\left(\overline{P}^{\rm SL}_{\Sigma \Sigma}-\overline{P}^{\rm SL}_{g g}\right)+R^{\rm SL}_{\Sigma g},
\nonumber\\
P^{\rm SL}_{g \Sigma}&=&\frac{1+\frac{C_A}{2C_{F}}\overline{C}_g^{\rm DL}}{1+\frac{C_A}{2C_{F}}C_g^{\rm
DL}}\overline{P}^{\rm SL}_{g \Sigma}+R^{\rm SL}_{g \Sigma},
\nonumber\\
P^{\rm SL}_{g g}&=&\overline{P}^{\rm SL}_{g g}+\frac{C_g^{\rm DL}-\overline{C}_g^{\rm DL}}{1+\frac{C_A}{2C_{F}}C_g^{\rm DL}}\overline{P}^{\rm SL}_{g \Sigma}
+\beta_0 a_s^2\frac{d}{da_s}\ln \frac{1+\frac{C_A}{2C_{F}}C_g^{\rm
DL}}{1+\frac{C_A}{2C_{F}}\overline{C}_g^{\rm DL}}+R^{\rm SL}_{g g},
\label{sc3}
\end{eqnarray}
with the definitions $P^{\rm SL}=(a_s/\omega)^2P^{[1]}(a_s/\omega^2)$, which are the complete SL contributions to the splitting functions, and 
$R^{\rm SL}=(a_s/\omega)^2R^{[1]}(a_s/\omega^2)$. 
Although our results for the SL contributions to the splitting functions in
Eq.\ (\ref{sc3}) depend on the
unknown components of the matrix $R^{\rm SL}$, it is clear from its form in Eq.\ (\ref{defofR}) that two of the four degrees of freedom
of the matrix $P^{\rm SL}$ are completely fixed. For example, these can be taken as any two of 
$P^{\rm SL}_{\Sigma \Sigma}-(2C_F/C_A)P^{\rm SL}_{g\Sigma}$,
$P^{\rm SL}_{gg}+(2C_F/C_A)P^{\rm SL}_{g\Sigma}$, and
the trace $P^{\rm SL}_{\Sigma \Sigma}+P^{\rm SL}_{g g}$.

We note that, interestingly, $R_{g\Sigma}^{\rm SL}=O(a_s^4)$ because, with this choice,
\begin{eqnarray}
P^{\rm SL}_{\Sigma \Sigma}&=&\frac{32 C_A C_F T_R n_f}{3\omega^4} a_s^3+O(a_s^4),
\nonumber\\
P^{\rm SL}_{g \Sigma}&=&\frac{2 T_R n_f}{3}a_s +\frac{16 C_A^2 T_R n_f}{3\omega^4} a_s^3 +O(a_s^4),
\nonumber\\
P^{\rm SL}_{g g}&=&-\frac{11C_A+4T_R n_f}{6}a_s
+\frac{2(11 C_A^2+4C_A T_R n_f-8 C_F T_R n_f)}{3\omega^2}a_s^2
\nonumber\\
&&{}-\frac{8(33C_A^3+12 C_A^2 T_R n_f-20 C_A C_F T_R n_f)}{3\omega^4}a_s^3 +O(a_s^4),
\label{checkPSLgS}
\end{eqnarray}
whose $O(a_s^2)$ terms are consistent with the results of Ref.\ \cite{Gluck:1992zx},
and whose $O(a_s^3)$ terms are consistent with the results of Refs.\ \cite{Moch:2007tx,Almasy:2011eq}.
Note also that $Z_{g\Sigma}^{[0]}=0$ (although we know from Appendix B of Ref.~\cite{Gaffney:1984yd} that $Z_{g \Sigma}^{[2]}\ne0$).
We will return to this point in the next section, where we investigate the effect of physical constraints on the remaining
undetermined degrees of freedom on our results.

\section{Physical constraints on the splitting functions}

In this section, we further constrain the SL contributions to the $\overline{\rm MS}$ splitting functions by exploiting some pysical properties of scheme-dependent quantities in physical schemes such as the $\overline{\rm MS}$ scheme.

According to Eq.\ (\ref{defofR}), $P^{\rm SL}_{\Sigma g}$ is the only component of the splitting function that is so far completely unconstrained,
even to $O(a_s^3)$,
since $R^{[1]}_{\Sigma g}$ also depends on the remaining three components of $Z^{[1]}$, and $Z_{\Sigma g}^{[1]}$ and $Z_{gg}^{[1]}$ are unknown.
Fortunately, this degree of freedom is fixed by assuming the absence of $\omega \to 0$ singularities
for all values of the factorization scale in $D_-$ defined in Eq.\ (\ref{DminusfromDSDg}). By inspection of the DGLAP equation in this basis,
%\be
%\frac{d}{d\ln \mu_f^2}D_{\pm}=\sum_{i=\pm}P_{\pm i}D_i,
%\ee
%this means 
it follws
that the splitting functions $P_{--}$ and $P_{-+}$ are found to be free of $\omega \to 0$ singularities,
i.e., neglecting all non-singular terms,
\be
P_{--}=P_{-+}=0.
\label{appassum}
\ee
This assumption is expected to be true to all orders. 
It is certainly true for the DL contributions to the timelike splitting functions, for the SL contributions in the MG scheme 
given in Eqs.\ (\ref{mgsplittingfunctions1})--(\ref{mgsplittingfunctions3}), and through NNLO \cite{Moch:2007tx,Almasy:2011eq},
as we verified in this paper.
Moreover, it is true through NNLO in the spacelike case \cite{Kotikov:1998qt,Illarionov:2004nw} and
holds for the leading and next-to-leading singularities to all orders in the
framework of Balitski-Fadin-Kuraev-Lipatov (BFKL) dynamics \cite{Fadin:1975cb,Kuraev:1976ge,Kuraev:1977fs,Balitsky:1978ic},
a fact that has been exploited in various approaches (see, for example, the recent papers
\cite{Ciafaloni:2007gf,Altarelli:2008aj} and the references cited therein).
We note that the timelike splitting functions share a number of simple properties with their spacelike counterparts,
e.g.\ the LO splitting functions are the same, and the diagonal splitting functions to all orders grow like $\ln \omega$ as $\omega \to \infty$.

Using the relations between the two bases in Eq.\ (\ref{1.7}), Eq.\ (\ref{appassum}) implies that
\begin{eqnarray}
P_{g \Sigma}&=&-\frac{\alpha}{\epsilon}P_{\Sigma \Sigma},
\nonumber\\
P_{\Sigma g}&=&-\frac{\epsilon}{\alpha}P_{g g}.\label{tolja3}
\label{properties}
\end{eqnarray}
where, through the SL level, which is all we need,
\be
-\frac{\epsilon}{\alpha}=\frac{2C_F}{C_A}\left[1+\omega\left(
\frac{1}{6}+\frac{1}{3}\frac{n_f T_R}{C_A}
-\frac{2}{3}\frac{C_F n_f T_R}{C_A^2}\right)\right],
\ee
which, at any order $k$, relates the two most singular terms in the off-diagonal splitting functions
$P_{\Sigma g}^{(k)}$ and $P_{g\Sigma }^{(k)}$ with those in the diagonal splitting functions $P_{\Sigma \Sigma}^{(k)}$ and $P_{gg}^{(k)}$.
Using Eq.\ (\ref{properties}), the assumption in Eq.\ (\ref{appassum}) implies that
\be
R^{\rm SL}_{g \Sigma}-\frac{C_A}{2C_F}R^{\rm SL}_{\Sigma \Sigma}=0,
\ee
which is already satisfied by the form in Eq.\ (\ref{defofR}), and
\be  
R^{\rm SL}_{\Sigma g}-\frac{2C_F}{C_A}R^{\rm SL}_{g g}=
\frac{4}{3}a_sC_F [C_A^2+2 n_f T_R (C_A-2 C_F)]
\left[\frac{C_g^{\rm DL}(C_A \overline{C}_g^{\rm DL}+C_F)}
{(C_A \overline{C}_g^{\rm DL}+2C_F)(C_A C_g^{\rm DL} +2C_F)}
\right]^2,
\label{rest}
\ee
which turns out to fix the following combination of $Z^{[1]}$ components:
\be
R^{\rm SL}_{\Sigma g}-\frac{2C_F}{C_A}R^{\rm SL}_{g g}
=-P_{gg}^{[0]}\left[
\frac{2C_F}{C_A}Z_{\Sigma\Sigma}^{[1]}
+Z_{\Sigma g}^{[1]}
-\left(\frac{2C_F}{C_A}\right)^2Z_{g\Sigma}^{[1]}
-\frac{2C_F}{C_A}Z_{g g}^{[1]}\right]\left(\frac{a_s}{\omega}\right)^2.
\ee

We can now write Eq.\ (\ref{defofR}) in the form
\be
R^{[1]}=\left(\begin{array}{cc}
0 &\frac{4}{3}a_sC_F [C_A^2+2 n_f T_R (C_A-2 C_F)]
\left[\frac{C_g^{\rm DL}(C_A \overline{C}_g^{\rm DL}+C_F)}
{(C_A \overline{C}_g^{\rm DL}+2C_F)(C_A C_g^{\rm DL} +2C_F)}
\right]^2
\\
0 & 0
\end{array}\right)+Z_{g\Sigma}^{[1]}\left(
\begin{array}{cc}
\frac{2C_F}{C_A} & -\left(\frac{2C_F}{C_A}\right)^2 \\
1 &  -\frac{2C_F}{C_A}
\end{array}
\right)\frac{P_{gg}^{[0]}}{Z_{gg}^{[0]}},
\label{finresforR1}
\ee
which shows that $Z_{g\Sigma}^{[1]}$ does not affect the evolution of the combination $(2C_F/C_A)D_g-D_\Sigma$,
reducing the dependence of the evolution on the unknown quantity $Z_{g\Sigma}^{[1]}$.

Note that Eq.\ (\ref{appassum}) (or, equivalently, Eq.\ (\ref{properties})) implies that the determinant of $P$ vanishes. 
Since the results in Eq.\ (\ref{sc3}) with $R^{[1]}$ given by Eq.\ (\ref{finresforR1}) imply that the trace is non-zero,
this means that one of the eigenvalues is zero and that the other one
coincides with the trace.

In order to complete the check of Eq.\ (\ref{checkPSLgS})
against the FO results in the literature, we need to consider the remaining
splitting function, for which we find
\be
P^{\rm SL}_{\Sigma g}=-3C_F a_s +\frac{12 C_A C_F}{\omega^2} a_s^2-
16\frac{(29 C_A^2 C_F+4C_F T_R n_f(C_A-C_F))}{3\omega^4} a_s^3 +O(a_s^4).
\ee
Here again, the $O(a_s^2)$ terms are in agreement with the results of Ref.\
\cite{Gluck:1992zx}, while the $O(a_s^3)$ terms are in agreement with the
results of Ref.\ \cite{Almasy:2011eq}.

We find that our resummed results exhibit the following $\omega\rightarrow 0$
behaviour:
\begin{eqnarray}
P_{\Sigma \Sigma}^{\rm SL}&=&\frac{4 C_F T_R n_f}{3 C_A^{3/4}} \frac{a_s^{5/4}}{\sqrt{\omega}}
-\frac{4C_F n_f T_R}{3C_A}a_s
+Z_{g\Sigma}^{[1]}\frac{P_{gg}^{[0]}}{Z_{gg}^{[0]}}\frac{2C_F}{C_A}
\left(\frac{a_s}{\omega}\right)^2
+O(\sqrt{\omega}),
\nonumber\\
P_{\Sigma g}^{\rm SL}&=& -\frac{8 C_F^2 T_R n_f}{3 C_A^{7/4}} \frac{a_s^{5/4}}{\sqrt{\omega}}
-\frac{C_F(11C_A^2 +4C_A n_f T_R-24 C_F n_f T_R)}{6C_A^2}a_s
-Z_{g\Sigma}^{[1]}\frac{P_{gg}^{[0]}}{Z_{gg}^{[0]}}\left(\frac{2C_F}{C_A}\right)^2\left(\frac{a_s}{\omega}\right)^2
+O(\sqrt{\omega}),
\nonumber\\
P_{g\Sigma}^{\rm SL}&=&\frac{2T_R n_f C_A^{1/4}}{3}
\frac{a_s^{5/4}}{\sqrt{\omega}}
+Z_{g\Sigma}^{[1]}\frac{P_{gg}^{[0]}}{Z_{gg}^{[0]}}\left(\frac{a_s}{\omega}\right)^2
+O(\sqrt{\omega}),
\nonumber\\
P_{gg}^{\rm SL}&=&-\frac{4 C_F T_R n_f}{3 C_A^{3/4}} \frac{a_s^{5/4}}{\sqrt{\omega}}
-\frac{11C_A^2+4C_A n_f T_R-8C_F n_f T_R}{12C_A}a_s
-Z_{g\Sigma}^{[1]}\frac{P_{gg}^{[0]}}{Z_{gg}^{[0]}}\frac{2C_F}{C_A}
\left(\frac{a_s}{\omega}\right)^2
+O(\sqrt{\omega}).
\end{eqnarray}
These limits imply the following nontrivial relation among the SL contributions to the $\overline{\rm MS}$ splitting functions: 
\be
\left[\frac{2 C_F}{C_A}P^{\rm SL}_{\Sigma \Sigma}
+P^{\rm SL}_{\Sigma g} 
-\left(\frac{2 C_F}{C_A}\right)^2 P^{\rm SL}_{g \Sigma}
-\frac{2 C_F}{C_A} P^{\rm SL}_{g g} \right]_{\omega=0}=0. \label{susy}
\ee
Equation (\ref{susy}) is also obeyed by the SL contributions in the MG scheme
\cite{Mueller:1983cq},
which can be checked using Eqs.\ (\ref{mgsplittingfunctions1}) and (\ref{mgsplittingfunctions3}),
and by the DL contributions, which are the same in both schemes.

It is interesting to observe that Eq.\ (\ref{susy}) is also true for all values of $\omega$ when the choice
$C_A=C_F=n_f$ is made, which corresponds to an $\mathcal{N}=1$ supersymmetric theory. 
Supersymmetry relations like the one in Eq.\ (\ref{susy}) were
first introduced in Refs.\ \cite{Dokshitzer:1977sg,Dokshitzer:1978hw} at one loop, 
then discussed at two loops in Refs.\ \cite{Furmanski:1980cm,Antoniadis:1981zv}
and, very recently, at three loops in Ref.\ \cite{Almasy:2011eq}.
Accidentally, as shown in Ref.\ \cite{Mueller:1983cq} in the timelike case,
Eq.\ (\ref{susy}) also reflects the fact that
an observable like the multiplicity ratio in quark and gluon jets is scheme independent. 

As we have seen, the only undetermined quantity appearing in our formulae, $R_{g\Sigma}^{\rm SL}$, is not constrained by any physical conditions.
In addition we noted that $R_{g\Sigma}^{\rm SL}=O(a_s^4)$.
This suggests that $R_{g\Sigma}^{\rm SL}$ is an artefact of the $\overline{\rm MS}$ scheme.

\section{Conclusions}
\label{conc}

In this paper, we presented the SL contributions to the $\overline{\rm MS}$ splitting functions.
Two of the degrees of freedom in the flavour-singlet matrix were determined
from the SL contributions to the MG splitting functions and the DL contributions to the $e^+ e^-$ coefficient 
functions in both the $\overline{\rm MS}$ and MG schemes.
One of the remaining two degrees of freedom was constrained by using
certain non-singular properties of the flavour singlet matrix at small values of $\omega$, which have been 
investigated only in the spacelike case so far (see, e.g., Ref.~\cite{Buras:1979yt}). Nevertheless, both 
eigenvalues are determined analytically in closed form.
Our results are in agreement with very recent calculations of the splitting functions 
in the $\overline{\rm MS}$ scheme at NNLO \cite{Moch:2007tx,Almasy:2011eq}
in the FO approach, and also with general physical requirements such as supersymmetry.

Knowledge of the complete SL contributions to the splitting functions formally improves the theoretical description of the 
evolution of FFs at small values of $\omega$ and thus facilitates the extraction of FFs from experimental data at small values of $x$ in global fits.
To date, such global fits have been performed to NLO in the FO approach
\cite{Albino:2005me,Albino:2005mv,Albino:2008fy}.
Our calculation of the SL contributions can be incorporated into such fits using
the consistent approach of Ref.\ \cite{Albino:2005gd}, which, together with the DL contribution to the $e^+ e^-$ coefficient function determined in Ref.~\cite{Albino:2011si},
allows for a description of the experimental data from the largest to the smallest $x$ values.
We recall that the NLO splitting functions contain also sub-SLs (sSLs), namely
the SGLs for which $m=3$ in Eq.\ (\ref{SGLexpanofP}), proportional to
$a_s^2/\omega$, but the complete sSL contributions to the splitting functions
are unknown.
In the SGL+FO(+FO$\delta$) scheme defined in Ref.\ \cite{Albino:2005gd},
these sSLs are, therefore, simply subtracted at this logarithmic order of accuracy.
Alternatively, these unresummed sSLs can be replaced by a simple matrix of sSL
functions, which are non-singular as $\omega \to 0$, but whose FO expansions
start with the NLO sSLs. An example of such a matrix of functions is
$a_s \gamma A/(2C_A)$.

{\it Note added.}
After the completion of this work, there appeared a preprint \cite{Vogt:2011jv} 
containing an alternative calculation of the SL contribution to the $\overline{\rm MS}$ splitting functions,
with which we found agreement, confirming both our approach and the approach of that article.
We stress that our approach highlights the relation between the splitting functions in the $\overline{\rm MS}$ scheme and those in the MG scheme,
the latter scheme being important in the scheme independent ratio of gluon to quark jet rates \cite{Mueller:1983cq}.
We also showed explicitly that our results are consistent with physical constraints.
Finally, we obtained closed forms for the splitting functions (up to $R_{g\Sigma}^{\rm SL}$) which we explicitly used to verify
the physical result in Eq.\ (\ref{susy}).

\subsection*{Acknowledgments}

P.B.\ kindly thanks L.N.\ Lipatov for valuable discussions concerning
supersymmetry relations among splitting functions.
The work of A.V.K.\ was supported in part by the Russian Foundation for Basic
Research RFBR through Grant No.\ 11--02--01454--a.
This work was supported in part by the German Federal Ministry for Education
and Research BMBF through Grant No.\ 05~HT6GUA, by the German Research
Foundation DFG through the Collaborative Research Centre No.~676
{\it Particles, Strings and the Early Universe---The Structure of Matter and
Space Time}, and by the Helmholtz Association HGF through the Helmholtz
Alliance Ha~101 {\it Physics at the Terascale}.


\begin{thebibliography}{99}

%\cite{Gribov:1972ri}
\bibitem{Gribov:1972ri}
  V.~N.~Gribov and L.~N.~Lipatov,
  %``Deep inelastic e p scattering in perturbation theory,''
  Sov.\ J.\ Nucl.\ Phys.\  {\bf 15} (1972) 438
  [Yad.\ Fiz.\  {\bf 15} (1972) 781].
  %%CITATION = YAFIA,15,781;%%

%\cite{Gribov:1972rt}
\bibitem{Gribov:1972rt}
  V.~N.~Gribov and L.~N.~Lipatov,
  %``e+ e- pair annihilation and deep inelastic e p scattering in perturbation
  %theory,''
  Sov.\ J.\ Nucl.\ Phys.\  {\bf 15} (1972) 675
  [Yad.\ Fiz.\  {\bf 15} (1972) 1218].
  %%CITATION = YAFIA,15,1218;%%

%\cite{Altarelli:1977zs}
\bibitem{Altarelli:1977zs}
  G.~Altarelli and G.~Parisi,
  %``Asymptotic Freedom In Parton Language,''
  Nucl.\ Phys.\  B {\bf 126} (1977) 298.
  %%CITATION = NUPHA,B126,298;%%

%\cite{Dokshitzer:1977sg}
\bibitem{Dokshitzer:1977sg}
  Y.~L.~Dokshitzer,
  %``Calculation Of The Structure Functions For Deep Inelastic Scattering And E+
  %E- Annihilation By Perturbation Theory In Quantum Chromodynamics,''
  Sov.\ Phys.\ JETP {\bf 46} (1977) 641
  [Zh.\ Eksp.\ Teor.\ Fiz.\  {\bf 73} (1977) 1216].
  %%CITATION = ZETFA,73,1216;%%

%\cite{Albino:2005gd}
\bibitem{Albino:2005gd}
  S.~Albino, B.~A.~Kniehl, G.~Kramer and W.~Ochs,
  %``Resummation of Soft Gluon Logarithms in the DGLAP Evolution of
  %Fragmentation Functions,''
  Phys.\ Rev.\  D {\bf 73} (2006) 054020
  [arXiv:hep-ph/0510319].
  %%CITATION = PHRVA,D73,054020;%%

%\cite{Albino:2011si}
\bibitem{Albino:2011si}
  S.~Albino, P.~Bolzoni, B.~A.~Kniehl and A.~Kotikov,
  %``Fully double-logarithm-resummed cross sections,''
  Nucl.\ Phys.\  B {\bf 851} (2011) 86
  [arXiv:1104.3018 [hep-ph]].
  %%CITATION = NUPHA,B851,86;%%

%\cite{Albino:2011bf}
\bibitem{Albino:2011bf}
  S.~Albino, P.~Bolzoni, B.~A.~Kniehl and A.~Kotikov,
  %``Timelike small x Resummation for Fragmentation Functions,''
  arXiv:1107.1142 [hep-ph].
  %%CITATION = ARXIV:1107.1142;%%

%\cite{Mueller:1983cq}
\bibitem{Mueller:1983cq}
  A.~H.~Mueller,
  %``Square Root Of Alpha (Q**2) Corrections To Particle Multiplicity Ratios In
  %Gluon And Quark Jets,''
  Nucl.\ Phys.\  B {\bf 241} (1984) 141.
  %%CITATION = NUPHA,B241,141;%%

%\cite{Albino:2005gg}
\bibitem{Albino:2005gg}
  S.~Albino, B.~A.~Kniehl, G.~Kramer and W.~Ochs,
  %``Generalizing the DGLAP evolution of fragmentation functions to the
  %smallest x values,''
  Phys.\ Rev.\ Lett.\  {\bf 95} (2005) 232002
  [arXiv:hep-ph/0503170].
  %%CITATION = PRLTA,95,232002;%%

%\cite{Albino:2005cq}
\bibitem{Albino:2005cq}
  S.~Albino, B.~A.~Kniehl, G.~Kramer and W.~Ochs,
  %``Unifying the fixed order evolution of fragmentation functions with the
  %modified leading logarithm approximation,''
  PoS {\bf HEP2005} (2006) 063
  [arXiv:hep-ph/0511228].
  %%CITATION = POSCI,HEP2005,063;%%

%\cite{Buras:1979yt}
\bibitem{Buras:1979yt}
  A.~J.~Buras,
  %``Asymptotic Freedom In Deep Inelastic Processes In The Leading Order And
  %Beyond,''
  Rev.\ Mod.\ Phys.\  {\bf 52} (1980) 199.
  %%CITATION = RMPHA,52,199;%%

%\cite{Kotikov:2002ab}
\bibitem{Kotikov:2002ab}
  A.~V.~Kotikov and L.~N.~Lipatov,
  %``DGLAP and BFKL evolution equations in the N=4 supersymmetric gauge
  %theory,''
  Nucl.\ Phys.\  B {\bf 661} (2003) 19
  [Erratum-ibid.\  B {\bf 685} (2004) 405]
  [arXiv:hep-ph/0208220].
  %%CITATION = NUPHA,B661,19;%%

%\cite{Albino:2008gy}
\bibitem{Albino:2008gy}
  S.~Albino,
  %``The hadronization of partons,''
  Rev.\ Mod.\ Phys.\  {\bf 82} (2010) 2489
  [arXiv:0810.4255 [hep-ph]].
  %%CITATION = RMPHA,82,2489;%%

%\cite{Mueller:1982cq}
\bibitem{Mueller:1982cq}
  A.~H.~Mueller,
  %``Multiplicity And Hadron Distributions In QCD Jets: Nonleading Terms,''
  Nucl.\ Phys.\  B {\bf 213} (1983) 85.
  %%CITATION = NUPHA,B213,85;%%

%\cite{Mueller:1983js}
\bibitem{Mueller:1983js}
  A.~H.~Mueller,
  %``Multiplicity And Hadron Distributions In QCD Jets. 2. A General Procedure
  %For All Nonleading Terms,''
  Nucl.\ Phys.\  B {\bf 228} (1983) 351.
  %%CITATION = NUPHA,B228,351;%%

%\cite{Gluck:1992zx}
\bibitem{Gluck:1992zx}
  M.~Gl\"uck, E.~Reya and A.~Vogt,
  %``Parton fragmentation into photons beyond the leading order,''
  Phys.\ Rev.\  D {\bf 48} (1993) 116
  [Erratum-ibid.\  D {\bf 51} (1995\ PHRVA,D51,1427.1995) 1427].
  %%CITATION = PHRVA,D51,1427;%%

%\cite{Moch:2007tx}
\bibitem{Moch:2007tx}
  S.~Moch and A.~Vogt,
  %``On Third-Order Timelike Splitting Functions and Top-Mediated Higgs Decay
  %into Hadrons,''
  Phys.\ Lett.\  B {\bf 659} (2008) 290
  [arXiv:0709.3899 [hep-ph]].
  %%CITATION = PHLTA,B659,290;%%

%\cite{Almasy:2011eq}
\bibitem{Almasy:2011eq}
  A.~A.~Almasy, A.~Vogt and S.~Moch,
  %``On the Next-to-Next-to-Leading Order Evolution of Flavour-Singlet
  %Fragmentation Functions,''
  Nucl.\ Phys.\  B {\bf 854} (2012) 133
  [arXiv:1107.2263 [hep-ph]].
  %%CITATION = NUPHA,B854,133;%%

%\cite{Gaffney:1984yd}
\bibitem{Gaffney:1984yd}
  J.~B.~Gaffney and A.~H.~Mueller,
  %``Alpha (Q**2) Corrections To Particle Multiplicity Ratios In Gluon And Quark
  %Jets,''
  Nucl.\ Phys.\  B {\bf 250} (1985) 109.
  %%CITATION = NUPHA,B250,109;%%

%\cite{Kotikov:1998qt}
\bibitem{Kotikov:1998qt}
  A.~V.~Kotikov and G.~Parente,
  %``Small x behaviour of parton distributions with soft initial conditions,''
  Nucl.\ Phys.\  B {\bf 549} (1999) 242
  [arXiv:hep-ph/9807249].
  %%CITATION = NUPHA,B549,242;%%

%\cite{Illarionov:2004nw}
\bibitem{Illarionov:2004nw}
  A.~Y.~Illarionov, A.~V.~Kotikov and G.~Parente Bermudez,
  %``Small x behavior of parton distributions. A study of higher twist
  %effects,''
  Phys.\ Part.\ Nucl.\  {\bf 39} (2008) 307
  [arXiv:hep-ph/0402173].
  %%CITATION = PPNUE,39,307;%%

%\cite{Fadin:1975cb}
\bibitem{Fadin:1975cb}
  V.~S.~Fadin, E.~A.~Kuraev and L.~N.~Lipatov,
  %``On The Pomeranchuk Singularity In Asymptotically Free Theories,''
  Phys.\ Lett.\  B {\bf 60} (1975) 50.
  %%CITATION = PHLTA,B60,50;%%

%\cite{Kuraev:1976ge}
\bibitem{Kuraev:1976ge}
  E.~A.~Kuraev, L.~N.~Lipatov and V.~S.~Fadin,
  %``Multi - Reggeon Processes In The Yang-Mills Theory,''
  Sov.\ Phys.\ JETP {\bf 44} (1976) 443
  [Zh.\ Eksp.\ Teor.\ Fiz.\  {\bf 71} (1976) 840].
  %%CITATION = ZETFA,71,840;%%

%\cite{Kuraev:1977fs}
\bibitem{Kuraev:1977fs}
  E.~A.~Kuraev, L.~N.~Lipatov and V.~S.~Fadin,
  %``The Pomeranchuk Singularity In Nonabelian Gauge Theories,''
  Sov.\ Phys.\ JETP {\bf 45} (1977) 199
  [Zh.\ Eksp.\ Teor.\ Fiz.\  {\bf 72} (1977) 377].
  %%CITATION = ZETFA,72,377;%%

%\cite{Balitsky:1978ic}
\bibitem{Balitsky:1978ic}
  I.~I.~Balitsky and L.~N.~Lipatov,
  %``The Pomeranchuk Singularity In Quantum Chromodynamics,''
  Sov.\ J.\ Nucl.\ Phys.\  {\bf 28} (1978) 822
  [Yad.\ Fiz.\  {\bf 28} (1978) 1597].
  %%CITATION = YAFIA,28,1597;%%

%\cite{Ciafaloni:2007gf}
\bibitem{Ciafaloni:2007gf}
  M.~Ciafaloni, D.~Colferai, G.~P.~Salam and A.~M.~Stasto,
  %``A matrix formulation for small-x singlet evolution,''
  JHEP {\bf 0708} (2007) 046
  [arXiv:0707.1453 [hep-ph]].
  %%CITATION = JHEPA,0708,046;%%

%\cite{Altarelli:2008aj}
\bibitem{Altarelli:2008aj}
  G.~Altarelli, R.~D.~Ball and S.~Forte,
  %``Small x Resummation with Quarks: Deep-Inelastic Scattering,''
  Nucl.\ Phys.\  B {\bf 799} (2008) 199
  [arXiv:0802.0032 [hep-ph]].
  %%CITATION = NUPHA,B799,199;%%

%\cite{Dokshitzer:1978hw}
\bibitem{Dokshitzer:1978hw}
  Y.~L.~Dokshitzer, D.~Diakonov and S.~I.~Troian,
  %``Hard Processes In Quantum Chromodynamics,''
  Phys.\ Rept.\  {\bf 58} (1980) 269.
  %%CITATION = PRPLC,58,269;%%

%\cite{Furmanski:1980cm}
\bibitem{Furmanski:1980cm}
  W.~Furmanski and R.~Petronzio,
  %``Singlet Parton Densities Beyond Leading Order,''
  Phys.\ Lett.\  B {\bf 97} (1980) 437.
  %%CITATION = PHLTA,B97,437;%%

%\cite{Antoniadis:1981zv}
\bibitem{Antoniadis:1981zv}
  I.~Antoniadis and E.~G.~Floratos,
  %``A Study Of A Possible Quark - Gluon Symmetry In QCD,''
  Nucl.\ Phys.\  B {\bf 191} (1981) 217.
  %%CITATION = NUPHA,B191,217;%%

%\cite{Albino:2005me}
\bibitem{Albino:2005me}
  S.~Albino, B.~A.~Kniehl and G.~Kramer,
  %``Fragmentation functions for light charged hadrons with complete quark
  %flavour separation,''
  Nucl.\ Phys.\  B {\bf 725}, 181 (2005)
  [arXiv:hep-ph/0502188].
  %%CITATION = NUPHA,B725,181;%%

%\cite{Albino:2005mv}
\bibitem{Albino:2005mv}
  S.~Albino, B.~A.~Kniehl and G.~Kramer,
  %``Fragmentation functions for K0(S) and Lambda with complete quark  flavour
  %separation,''
  Nucl.\ Phys.\  B {\bf 734}, 50 (2006)
  [arXiv:hep-ph/0510173].
  %%CITATION = NUPHA,B734,50;%%

%\cite{Albino:2008fy}
\bibitem{Albino:2008fy}
  S.~Albino, B.~A.~Kniehl and G.~Kramer,
  %``AKK Update: Improvements from New Theoretical Input and Experimental
  %Data,''
  Nucl.\ Phys.\  B {\bf 803}, 42 (2008)
  [arXiv:0803.2768 [hep-ph]].
  %%CITATION = NUPHA,B803,42;%%

%\cite{Vogt:2011jv}
\bibitem{Vogt:2011jv}
  A.~Vogt,
  %``Resummation of small-x double logarithms in QCD: semi-inclusive
  %electron-positron annihilation,''
  JHEP {\bf 1110} (2011) 025
  [arXiv:1108.2993 [hep-ph]].
  %%CITATION = JHEPA,1110,025;%%

\end{thebibliography}
\end{document}